# RECONSTRUCTING THE ANTIKYTHERA MECHANISM'S CENTRAL FRONT DIAL PARTS – DIVISION AND PLACEMENT OF THE ZODIAC DIAL RING

**Aristeidis Voulgaris**
*City of Thessaloniki, Directorate of Culture and Tourism
Thessaloniki, GR-54625, Greece.*
E-mail: arisvoulgaris@gmail.com

**Christophoros Mouratidis**
*Merchant Marine Academy of Syros, GR-84100, Greece.*
E-mail: christophoros.mouratidis@gmail.com

**and**

**Andreas Vossinakis**
*Thessaloniki Astronomy Club, Thessaloniki, GR-54646, Greece.*
E-mail: andreas.vossinakis@gmail.com

**Abstract:** In this paper we analyze, discuss and present the design of the Antikythera Mechanism's central front parts. Based on the aligned and visual images of the same scale of Fragment C front/back face and X-ray CT scans, we designed and reconstructed in bronze, the four independent parts comprising the central front dial. We then correlated the zodiac dial ring with 365 equal subdivisions/days and we investigated the number of days per astronomical season and per zodiac month. Then, we adopted a specific number of equal subdivisions/days per zodiac month and we engraved these on the bronze zodiac month ring. The different number of days per zodiac month created 12 unequal central angles on the zodiac dial ring and therefore the solar anomaly and the unequal time span of the astronomical seasons were well represented on the Antikythera Mechanism. In this way, the functionality of the central front dial of the Mechanism was achieved by adopting the minimum number of hypotheses.

## 1 INTRODUCTION

The Antikythera Mechanism is a complex geared device, which was constructed for time-measuring calculations and astronomical predictions based on the luni-solar cycle. The results were calculated by means of a large number of engaged gears and pointers which rotated on their corresponding scales. The Mechanism could calculate the (timed) position of the Sun on the Ecliptic, the phases of the Moon, solar and lunar eclipses,[1] the Metonic month (Anastasiou et al., 2016; Freeth et al., 2006), the hour when eclipses occurred[2] (Voulgaris et al., 2023b) and the starting date of the athletic Games (Freeth et al., 2008). The procedures of the Mechanism were presented versus time and were mostly based on the lunar synodic cycle, except for the Egyptian and zodiac dials, which were based on the solar cycle (see Table 1). The hypothesis that the planets' indication gearing on the Antikythera Mechanism presented the motion of the planets with rotated spheres is highly doubtful because it is not in accordance with the *Construction Protocol for a Research Quality Bronze Reconstruction of the Antikythera Mechanism*, presented in the Appendix (Section 8) at the end of this paper.

Today the Mechanism is partially preserved in seven relatively large fragments and 75 smaller ones. During its 2000 years underwater, the bronze material (an alloy of ~94% copper and ~6% tin, Price, 1974) of the Mechanism (density 8.87gr/cm$^3$) was transformed into atacamite (see Voulgaris et al., 2019b). Atacamite is a copper halide mineral with a significantly lower density (3.76gr/cm$^3$) than bronze (http://webmineral.com/data/Atacamite.shtml).

Owing to the significant change in the material density, as well as to corrosion and gravity, the Mechanism's parts suffered irregular alterations in shape, volume and position. With time, the fragile corroded pieces of the Mechanism started to collapse inside the Mechanism's rotten box.[3] An additional strong deformation of the Mechanism happened when in 1901 it was removed from the bottom of the sea near Antikythera Island: the abrupt dehydration of the Mechanism and the dry atmosphere further contributed to the irregular deformation, displacement, shrinkage and change of volume (dimensions and shape) of the Mechanism's parts (Voulgaris et al., 2019b), e.g. see the strong deformation of Fragment C shown in Figure 1E.

Table 1: The preserved dials of the Antikythera Mechanism and their corresponding units. The type of measurement for each scale is the same: Units of time (the units of the zodiac dial ring are analyzed in this work). The nature of the units in a measuring instrument defines the kind of instrument and what physical quantity it counts. Any measuring instrument has calibrated scales that present relevant or associated measurements.

| Measuring Dial | Antikythera Mechanism Scale | Scale Units/ Subdivisions | Type of Measurement |
|---|---|---|---|
| Egyptian dial ring | Egyptian Year of 365.0 Days | 1 day/subdivision | Time |
| Parapegma plates 1, 2 | 1 Year = 2 Plates = 4 Columns<br>1 Column = 1 Season = 3 months<br>Each parapegma event corresponds to a specific date = 1 day per index letter | Event/day and index letter/day | Time |
| Lunar phases sphere | The lunar phases are measured in days (e.g. Full Moon = $15^d$ Moon, not Moon at 180°) | Colors: black, white, half black/half white | Time |
| Metonic spiral | 19 Years/235 synodic months | 1 synodic month/cell | Time |
| Saros spiral | 18.03 Years/223 synodic months | 1 synodic month/cell | Time |
| Exeligmos dial | 54.09 = 3 × Saros (= 3 × 223 synodic months) | 18.03 years/⅓ sector | Time |
| Athletic Games dial | 4 years (= alternately 49/50 synodic months) | 1 year/quadrant | Time |
| Zodiac dial ring | Full circle corresponds to one solar tropical year | 360 degrees or 365 days | Space or time |

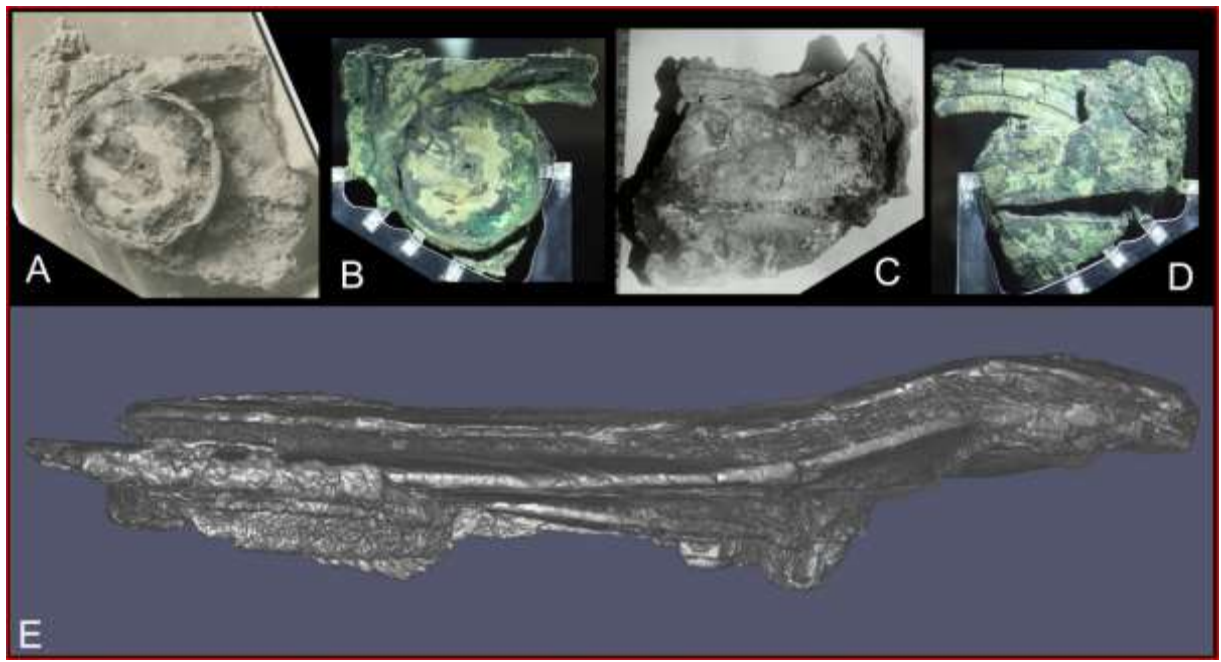

Figure 1(A) and (C): Fragment C (back and front faces) captured in 1903 by an anonymous photographer and included in Rehm (1906; cf. Rehm 1905). Around 1905, the two parapegma plates were better preserved than they currently are. (B and D): The current condition of Fragment C, back and front faces. Credits: National Archaeological Museum, Athens, first author, Copyright Hellenic Ministry of Culture & Sports/Archaeological Receipts Fund. (E): A 3-D reconstruction of Fragment C based on the AMRP X-Ray Raw Volume,[4] Top Coronal view of Fragment C. The permanent strong deformation of Fragment C (bent and twisted) is evident. X-ray Volume process by the authors, via REAL3D VolViCon software.

Therefore, all modern measurements are approximate, since these measurements are applied to the deformed parts—not to the original bronze prototype—and in many cases they are not representative (e.g. see the graphs in Voulgaris et al., 2022: Figure 8). More realistic approaches for the parts' dimensions can be deduced by taking into account the symmetry of the design and the ideal/correct positions for the best operation and functioning of the Mechanism's parts.

A large number of cracks in the material are visible to the naked eye, in the radiographs and in the computed tomography (CT) scans.

Today, the fragments are totally (or deeply) corroded inside (see Voulgaris et al., 2021: Figure 1) and their degree of X-ray absorption (darkening) is almost everywhere equal to the atacamite absorption (Ramsey, 2012; Voulgaris et al., 2018c; 2021; also, see Figure 1).

The front face of the Mechanism consists of three independent sectors: the two parapegma plates 1 and 2 (PP1-top and PP2-bottom) and the central front dial (Anastasiou et al., 2013; Bitsakis and Jones, 2016a). Today, the main part of the front face is partially preserved as Fragment C: the central front dial (~20%) and a part of parapegma plates 1 and

2 (see Figure 1).[5]

The central front dial of the Antikythera Mechanism is the main cadran (dial) of the Mechanism. This cadran gives important information for time measuring based on the lunar sidereal/synodic cycles and the solar tropical year: the synodic lunar cycle (via the lunar pointer which travels around the Ecliptic and is either aligned with or points opposite to the golden sphere, New Moon/Full Moon),[6] the sidereal cycle (when the lunar pointer points to the same position of a zodiac constellation), the phases of the Moon (via the Moon phases sphere located at the lunar cylinder's perimeter) and the solar tropical year/cycle (via the solar pointer of the ΧΡΥΣΟΥΝ ΣΦΑΙΡΙΟΝ, golden sphere,[7] which also travels along the Ecliptic).

The Ecliptic and the 12 zodiac constellations are represented on the zodiac month ring.

Encircling the zodiac dial ring, a second outer ring represents the Egyptian (year) calendar (Price, 1974). For several astronomical events, Ptolemy refers to their dates in the Metonic (also in Dionysius, see footnote 23) and the Egyptian calendars (Heiberg, 1898: 1.341, 1.342, 2.25, 2.28, 2.29, 2.32, 2.268, 2.419; Toomer, 1984). This practice must have been widespread during the Hellenistic era, since Greece, Egypt and Babylon were considered as one state and their administrations, policies, cultures and knowledge were interconnected (Voulgaris et al., 2023c).

## 2   THE DESIGN OF THE ANTIKYTHERA MECHANISM FRONT DIAL PLATE

### 2.1   Analysis of the Design of the Front Plate Parts

The central front plate is comprised of the two dial scale rings, the Egyptian and the zodiac, which are homocentric to each other, as well as to the lunar cylinder and b1 gear.

The Egyptian dial ring represents the Egyptian year (Price, 1974), which had a constant duration of 365.0 days/year, breaking down into 12 months × 30 days + 5 ΕΠΑΓΟΜΕΝΑΙ epagomenal days, dedicated to the birthdays of Osiris, Horus, Seth/Typhon, Isis and Nephthys (Maravelia, 2018b; Theodosiou and Danezis, 1995). The Egyptian calendar was ahead by one day every four solar tropical years or one month every 120 years, as Geminus mentions (Evans and Berggren, 2006; Manitius, 1880, viii, 18–24).

The zodiac dial ring on the Mechanism is directly correlated with the solar tropical year, in which the Sun returns to its starting position (1st day of Capricorn–Winter Solstice; Voulgaris et al., 2023a)[8] in 365.25 days (van der Waerden, 1953). The zodiac dial ring is divided into 12 zodiac sectors, one for each of the twelve Ecliptic constellations. The pointer of the golden sphere (ΗΛΙΟΥ ΑΚΤΙΝ; Bitsakis and Jones, 2016a) traverses each of the zodiac sectors, representing the motion of the Sun as it is projected on the sky. The Sun traverses each zodiac sign of 30° over an unequal number of days due to the solar anomaly (discussed in Section 2.2).

For the reconstruction of the square front plate (design, parts and scales) the authors used their own photographs of Fragment C. The photographs were captured by an optomechanical system dedicated to this project and designed and constructed by the first author (Voulgaris et al., 2018a). Photographs taken by K. Xenikakis and Polynomial Texture Maps (PTMs) also made in 2004 were used to test the results. Images of the front and back faces of the fragments were captured at the same scale in pixels/mm (*ibid.*). All preserved parts of the front dial are relatively thin (~2mm), so simultaneous inspection of the front and back faces of Fragment C can provide us with information about their geometrical characteristics and positioning. Aligned photographs of the fragment's front and back faces allow their juxtapositioning by mirroring one of them.

According to Figure 2, the central front dial of the Antikythera Mechanism consists of four totally independent parts (see Voulgaris et al., 2018a):

(1) The square, or almost square, front plate, with a width of ~174mm according to Voulgaris et al. (2019b: Table 1; Allen et al., 2016);[9]
(2) The bearing base ring with 365 holes in a circular distribution (at present about 80 are preserved, Evans et al., 2010), which is stabilized on the back face of the square plate;
(3) The free-to-rotate Egyptian dial ring, which is placed above the bearing base ring; and
(4) The zodiac dial ring, which is also free to rotate clockwise and counterclockwise (similar results were obtained by applying the same process to the 2004 photographs obtained by K. Xenikakis and the PTM images).

Figure 3 shows the reconstruction of the central front dial plate parts of Fragment C in bronze, by applying the symmetry options to the design (Voulgaris et al., 2022).

By applying a middle-horizontal/vertical/

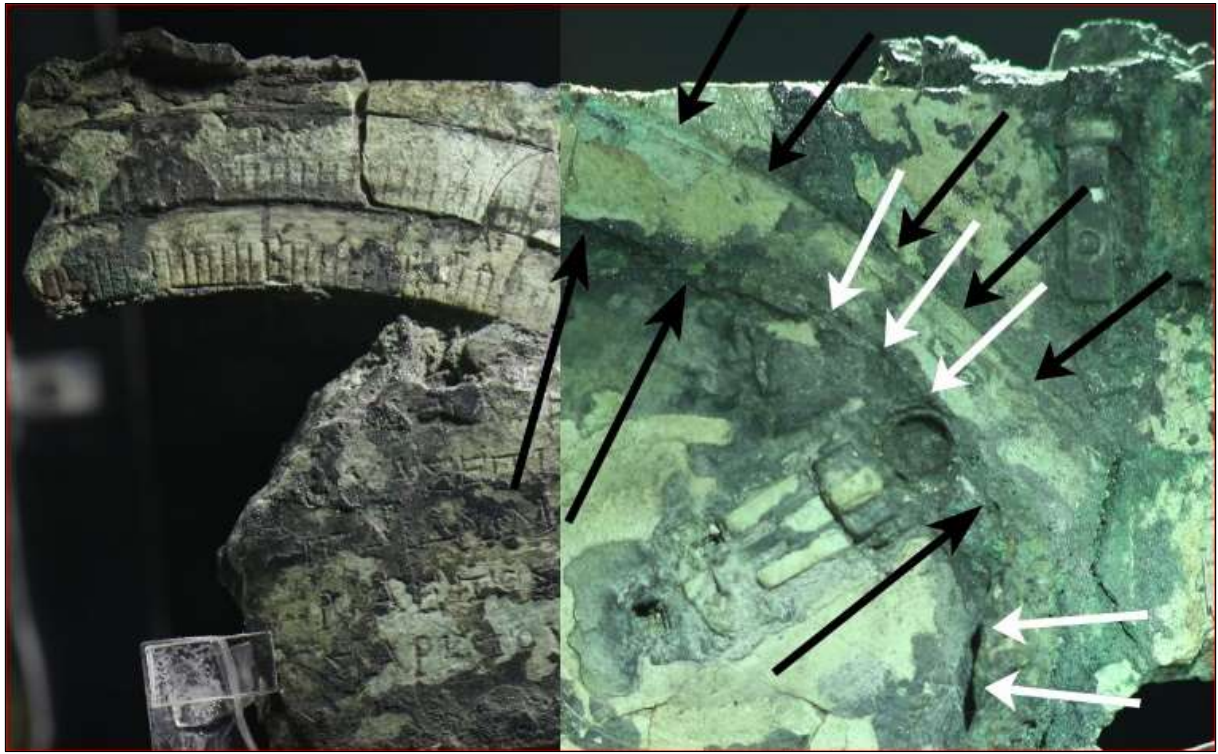

Figure 2: Composite image of same-scale aligned front/back faces of Fragment C. On the front face of Fragment C, the Egyptian and the zodiac dial rings are at the same level on the square front plate. The bearing base (black arrows) is visible on the back face of the fragment (see also Voulgaris et al., 2018a). The bearing base is partially covered by the lunar cylinder (its round contour is shown by the white arrows). The color balance of the right-hand photograph was deliberately altered to increase contrast and improve visibility compared to the left-hand photograph. Images processed by the authors. The sliding catch is visible on the top-right edge of the Fragment (Bitsakis and Jones, 2016a). (Credits: National Archaeological Museum, Athens, first author, Copyright Hellenic Ministry of Culture & Sports/Archaeological Receipts Fund).

mirror/central symmetry, it was deduced that this plate must have been almost, or exactly, square in shape (see also Allen et al., 2016).

It is difficult to discern the bearing base ring as a separate part on the radiographs and the CT scans, since it is corroded, worn out and stuck to the Egyptian dial ring, with the two of them essentially forming one piece. However, in the X-ray imaging the two stuck plates can be seen as independent, since there is a thin layer of air between them and air exhibits very weak absorption in X-rays (Day and Taylor, 1948; Voulgaris et al., 2018c); otherwise these two plates merely appear to be one thicker plate.

In the X-ray and CT scans of Fragment C about 80 of the ~0.7mm–0.8mm diameter 365 holes are visible.[10] The holes are clogged with petrified silt, but since the X-ray absorption for silt is very low relative to atacamite the holes are easily visible on the radiographs and the CT scans. As Figure 4 illustrates, the authors also captured some of the preserved holes on photographs, using a specifically positioned digital single lens reflex (DSLR) camera.

The square front plate as well as the two parapegma plates are based on the internal wooden casement (Voulgaris et al., 2019b). Since the square front plate is detachable via the four sliding catches, the two parapegma plates must have been stabilized on the internal wooden casement, probably with pins (see Figure 5).

The precise positioning of the square front plate was very critical, so that the rotational centers of the lunar cylinder and the golden sphere coincided precisely with the geometrical centers of the Egyptian and zodiac dial rings (preventing errors of eccentricity).

For this reason, the ancient craftsman made four small square holes on the corners of the square front plate (only one is preserved at present), which were the holes for the locking position of the plate. Originally square pin-drivers in these square holes were attached to the internal wooden casement (see Figure 5(B) and (C)).

On Fragment F, which is a part of the Saros spiral, also present is a corner of a plate with a thumb button, a sliding catch and its driver with a Ω-shape. This part is stuck above

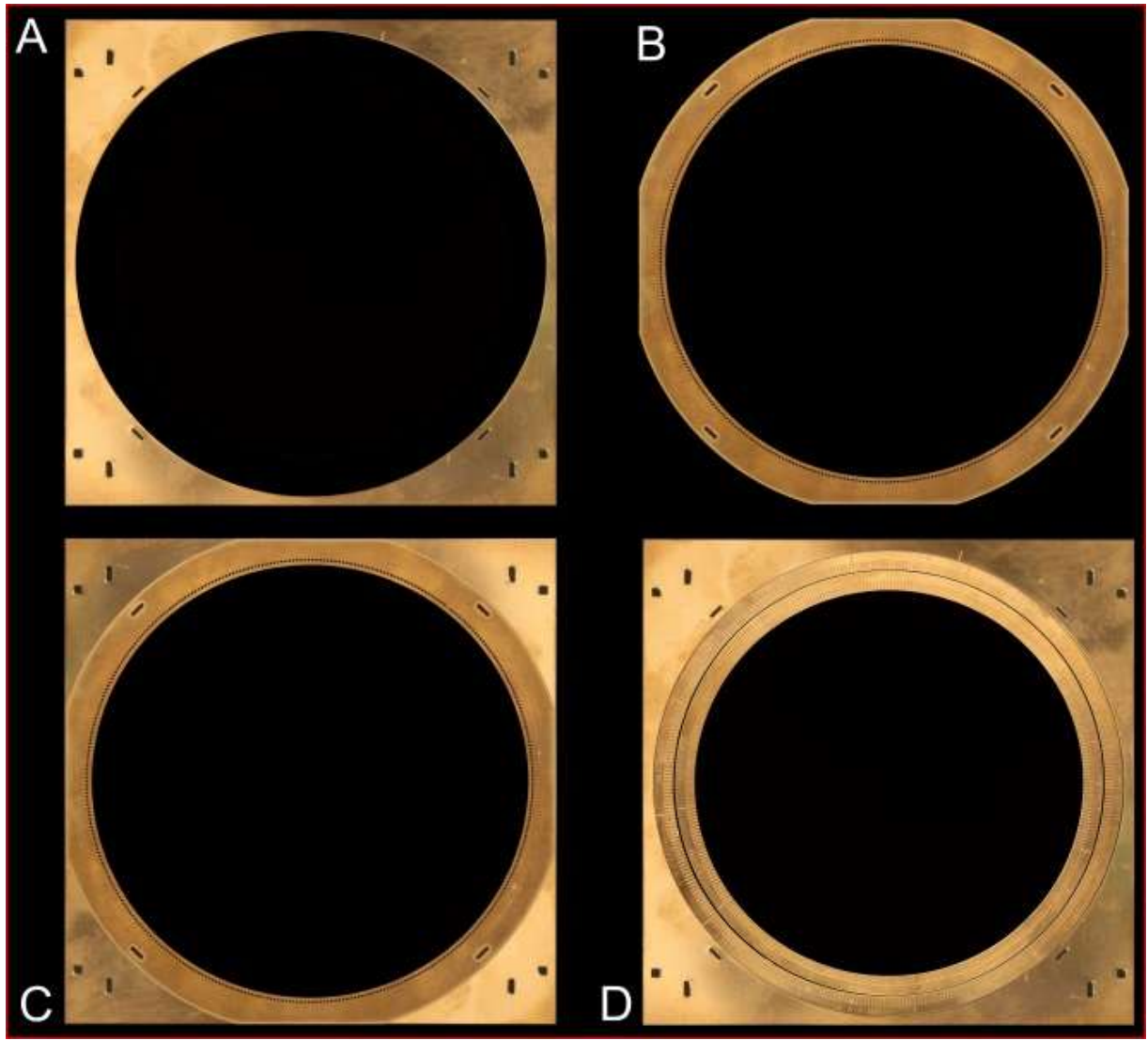

Figure 3: The four parts of the central front dial plate, reconstructed of bronze (94%Cu + 6%Sn). (A) The square front plate of the Mechanism. The ancient craftsman adapted four sliding catches on the small oblong holes, so that he could easily detach the square front plate to have a direct look at the moving parts inside (i.e. the gears and axles). (B) The bearing base ring (fourth part/third ring). The outer perimeter/arc limits of the bearing base ring are cut to match the square shape of the central square plate. On the bearing base ring, the ancient craftsman drilled 365 holes in a circular distribution. For the drilling of the holes on the reconstructed model, common electric tools were used, not computer-controlled machines. (C) The bearing base ring is stabilized on the back face of the central square plate. In this way two uneven levels are created. (D) The front dial of the Antikythera Mechanism: The central square front plate with the two independent and homocentric rings of the Egyptian and zodiac calendars. The bearing base ring is beneath the Egyptian dial ring. Bronze parts construction (photographs: the authors).

the Saros spiral parts and is surrounded by petrified silt (see Figure 6).

The strong similarity to the design of the preserved corner of the square front plate of Fragment C is evident in Figure 2 (see also Figure 3.3 in Bitsakis and Jones, 2016b: 76–77). According to its geometry, this piece must have been at the top left corner of the square front plate or at the bottom right (due to its proximity to the preserved Fragment F; note that due to corrosion, some components or parts of it collapsed inside the Mechanism's box; see the current position of the lunar disk in Figure 1B).[11]

## 2.2 The Zodiac Dial Ring of the Antikythera Mechanism and its Units

The zodiac dial ring of the Antikythera Mechanism represents the Ecliptic, which is the line in the middle of the 12° wide zodiac belt, as Geminus mentions in V.51–53 (see also van der Waerden, 1953).[12] In one full turn of the golden sphere–Sun, the Sun pointer travels through the 12 zodiac signs which is equal to one tropical year (and one full turn of the lunar cylinder around the Ecliptic equals one sidereal month of 27.321 days).

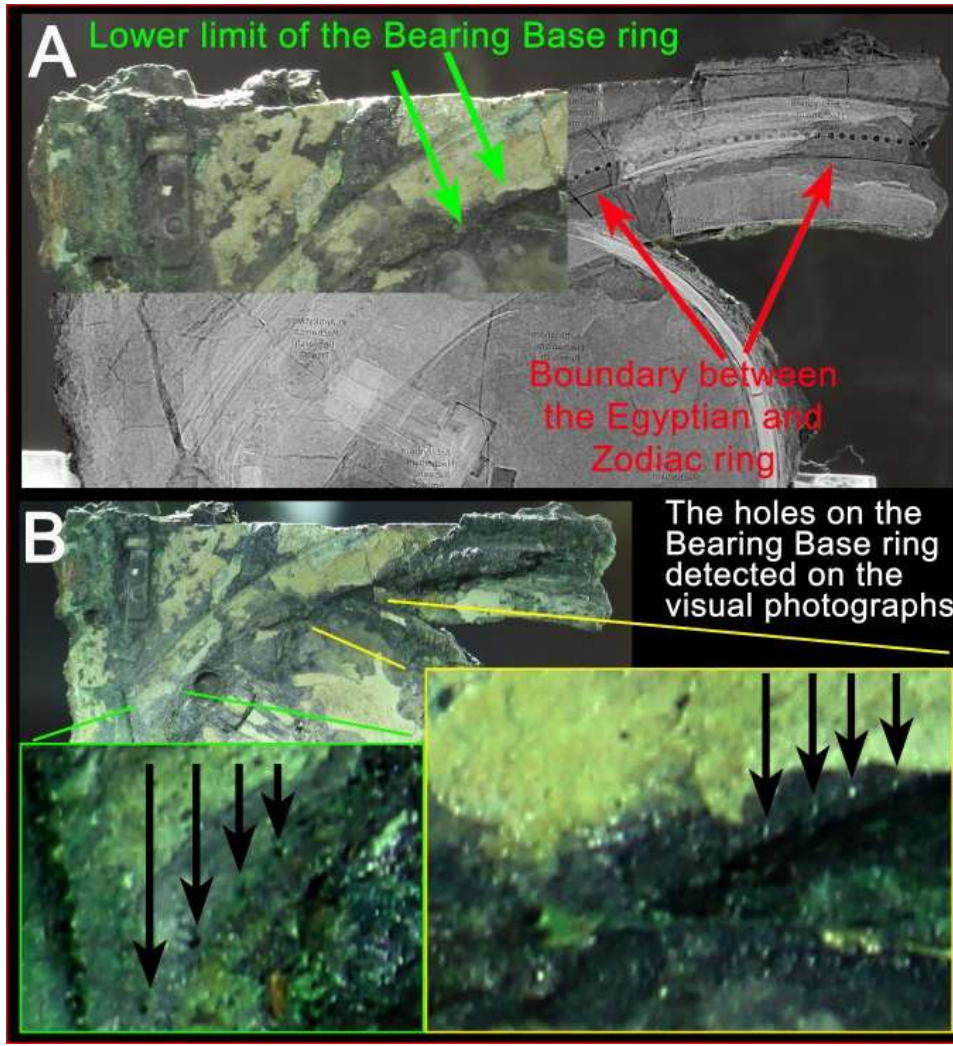

Figure 4(A): A composite image of a photograph and the AMRP radiograph, shown to scale (the aligning of the two images was based on the fragment's boundary and on the pins of the sliding catch). The limit between the Egyptian and zodiac dial rings is visible in the radiographs. The lower limit of the bearing base ring is clearly visible in the photograph. (B): The small holes in the bearing base ring were also detected in the photographs of Fragment C2. As the holes are filled with petrified silt they are visible as small pits (indicated by the black arrows). (Image by the first author. Credits: National Archaeological Museum, Athens, first autor, Copyright Hellenic Ministry of Culture & Sports/ Archaeological Receipts Fund).

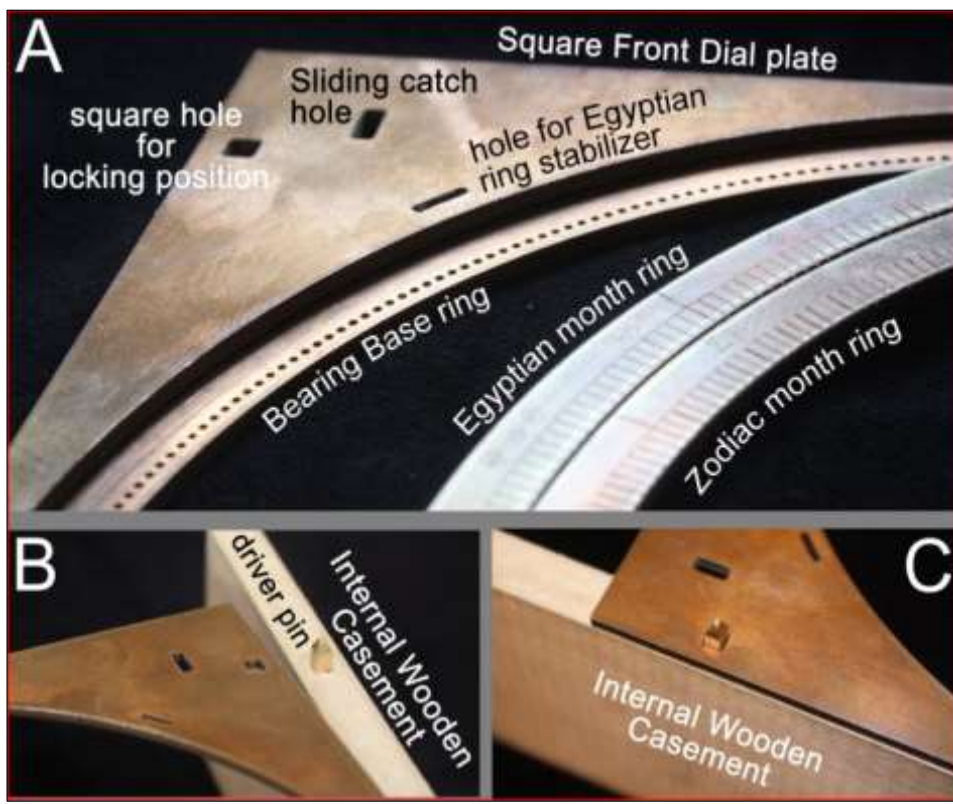

Figure 5(A): The bearing base ring is stabilized on the back face of the central square plate. In this way, the Egyptian dial ring and the zodiac month rings are located at the same level as the central square plate (see also Voulgaris et al., 2018a: Figure 4). (B): The square driver pin (missing and represented here) is fixed on the internal wooden casement of the Mechanism (Voulgaris et al., 2019b). (C) The driver pin enters its corresponding square locking hole, so that the central square plate is precisely positioned on the internal wooden casement. This method guarantees a high degree of accuracy when repositioning the plate.

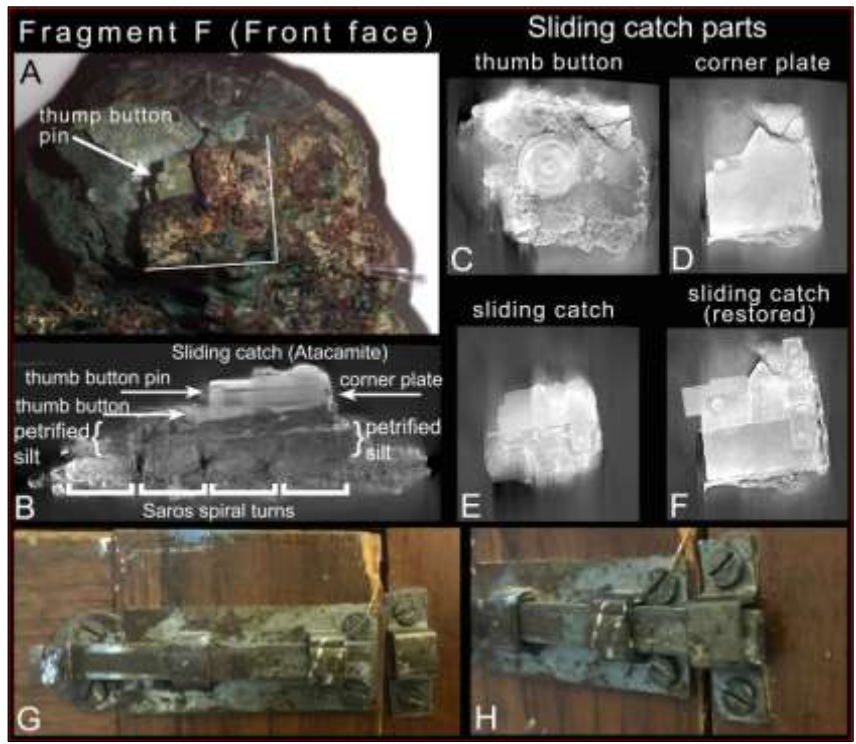

Figure 6(A): The front face of Fragment F on which the eclipse events are engraved (only visible in the CT scans). A corner of a plate (noted by the white lines) with a sliding catch is located about 5–7 mm above the fragment's surface (see Note 10). Fragment F and this plate were mostly surrounded by calcareous encrustation, even after their first (?) cleaning (around 1903–1905), created by deposits of marine carbonates, salts, dead microorganisms, etc. (Voulgaris et al., 2019b). The pin that connects the thumb button and the sliding catch is visible (image by the first author. Credits: National Archaeological Museum, Athens, first author, Copyright Hellenic Ministry of Culture & Sports/Archaeological Receipts Fund). (B): CT scan coronal view of Fragment F, specifically the area of the sliding catch (the slice aligned to the corner plate surface). (C): The thumb button of the sliding catch. (D): The plate's surface. (E): The sliding catch with its Ω-shaped driver, which is stabilized on the plate with pins. (F): A digital restoration of the sliding catch (CT scans processed by the authors via REAL3D VolViCon). (G) and (H): A 'modern' sliding catch from a 70-year old wooden storage cupboard (images: the authors).

Price (1974) suggested the subdivision of the zodiac dial ring into 360 degrees (see also Bitsakis and Jones, 2016a; Carman and Evans, 2014; Freeth et al., 2006; Wright, 2006). Price's assumption was based on measuring the (well) preserved subdivisions of the complete month of ΧΗΛΑΙ (Claws–Libra) and ΣΚΟΡΠΙΟΣ (Scorpion) which both have 30 subdivisions (see also Bitsakis and Jones, 2016a), and he also argued that all the other months were also divided into 30 subdivisions, i.e. 360 subdivisions/degrees in the ring. In order to represent the solar anomaly, these 360 subdivisions/degrees should be engraved at unequal distances from each other: larger distances for the months Gemini/Cancer (slower angular velocity) and shorter distances for the months Sagittarius/Capricorn (faster angular velocity)—see Evans et al. (2010).

It seems that the Antikythera Mechanism was a device that measured time (see Table 1), and this paper examines the scenario that the zodiac dial ring was divided into units of time, in 365 subdivisions-days (instead of arc degrees, which are units for angle/space measurement). Thus, star risings and settings on the parapegma were measured in days (units of time) and they were classified for each zodiac month (see Rehm, 1913; 1941; 1949).

If the 365.25 days were represented by dividing the zodiac dial ring in (rounded) 365 equal subdivisions/days,[13] then each zodiac month did not have an equal number of days, since this varied according to the total days the Sun needed to traverse each zodiac sign. This value depended on the solar anomaly phase, i.e. perihelion/aphelion position (see also Voulgaris et al., 2018a). This division was in accordance with the Mechanism's operation and measurements, eliminating the need for any complex assumptions (see the Appendix at the end of this paper).

Additionally, it was faster and easier to construct and divide the zodiac dial ring into 365 parts than to produce the Egyptian dial ring and the bearing base ring, both of which had to be divided into 365 parts but also have holes drilled correspondingly. That said, the division of these two rings could be done simultaneously by an ancient craftsman.

### 2.3 Dividing the Zodiac Dial Ring into 365 Equal Subdivisions/Days

The Egyptian and the zodiac dial rings have equal design, were concentric and their common centers coincided with the $b_{in}/b_{out}$ axis (which was also the center of the Mechanism's lunar disk), which was where the Earth was located.

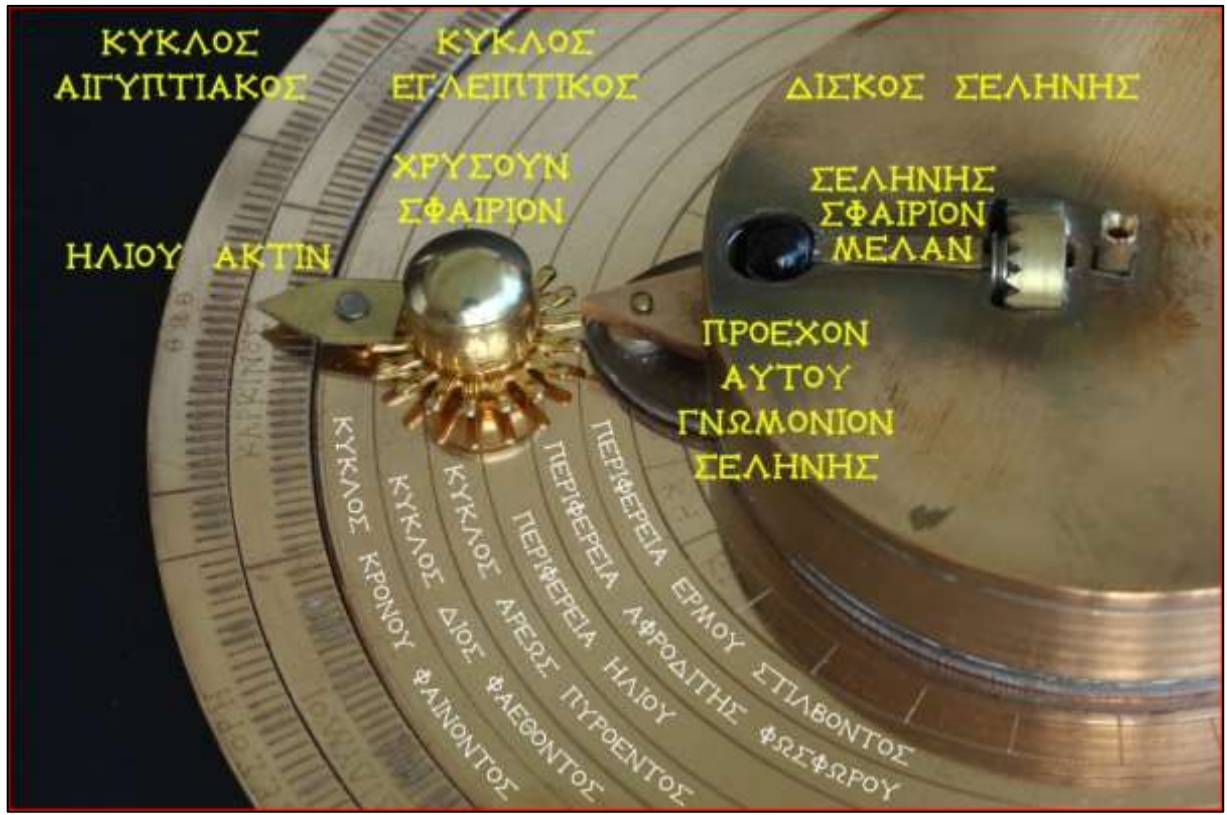

Figure 7: A close-up on the central front area of the authors' bronze reconstructed functioning model of the Antikythera Mechanism (here only the relevant parts are assembled): the lunar disc—ΔΙΣΚΟΣ ΣΕΛΗΝΗΣ (the ideal and proper input to the device; Voulgaris et al., 2018b; 2022; 2024; Roumeliotis, 2018); its lunar pointer—ΠΡΟΕΧΟΝ ΑΥΤΟΥ ΓΝΩΜΟΝΙΟΝ Σ[ΕΛΗΝΗΣ; the golden sphere–Sun—ΧΡΥΣΟΥΝ ΣΦΑΙΡΙΟΝ; its solar pointer Sun ray—ΗΛΙΟΥ ΑΚΤΙΝ (Bitsakis and Jones, 2016a); the zodiac month ring—ΚΥΚΛΟΣ ΕΓΛΕΙΠΤΙΚΟΣ; and the Egyptian dial ring—ΚΥΚΛΟΣ ΑΙΓΥΠΤΙΑΚΟΣ (as defined by the authors). If the two rings (which in this photograph are randomly placed) are divided into days, then any date on the zodiac month ring can correspond to a date in the Egyptian calendar (and *vice versa*). In the photograph (by the authors), the lunar pointer points at the golden sphere–Sun, New Moon phase and therefore the lunar phases sphere is totally black. The solar pointer points at the zodiac day subdivision of 22 Cancer, which corresponds to the 16[th] day of Thoth. The position of the two rings changes by one subdivision every four rotations of the golden sphere. The lost cover of the b1-gear is also presented, in which six circles are engraved, presenting the orbit of each of the five planets and the Sun, according to the preserved phrase on the back cover Inscription (Bitsakis and Jones, 2016b). ΚΡΟΝΟΥ ΦΑ[ΙΝΟΝΤΟΣ ΚΥΚΛΟΣ—orbit of Kronos–Phainon–Saturn; ΠΕΡΙΦΕΡΕΙΑ—(circumference) for Mercury, Venus and the Sun; ΚΥΚΛΟΣ—(circle) for Mars, Jupiter and Saturn (see the Appendix at the end of this paper).

The scale of the Egyptian dial ring represents the duration of the Egyptian year: 12 months of 30 days plus 5 days named *ΕΠΑΓΟΜΕΝΑΙ* epagomenal, in total 365.0 days.

The second dial scale with a common center to the Egyptian dial ring, the zodiac dial ring could also be a time-measuring scale, which represents the solar tropical year. Under this assumption, the two concentric and successive Egyptian and zodiac scales have equal design and represent the concept of the year's duration in units of time for these two different systems. Both systems are directly related to the timed motion of the Sun (while the lunar year has no relation to this operation). The time difference between the Egyptian year (= 365.0$^d$) and one full turn of the golden sphere–Sun (= 365.25$^d$), becomes one day every four solar tropical years, as Geminus also mentions. This procedure can be mechanically represented on the Mechanism by turning the Egyptian dial ring by one sub-division-day counterclockwise every four full turns of the golden sphere.

The division of the two rings into 365 days creates a visualization of this time difference between the two calendars. For a specific year, the user can match any date between the two rings e.g. which Egyptian date corresponds to the Summer Solstice (1[st] Cancer) or to a specific star event, or which zodiac date corresponds to the 17[th] day of Hathyr, which is the starting date of the Isia Feast (Jones 2000; Voulgaris et al., 2023a; 2023c); see Figure 7.

Additionally, the parapegma events occur on specific dates (not in degrees of Ecliptic long-itude), and the parapegma units-days would be in better agreement if the zodiac subdivisions were in days, since the index letters of the events are engraved on some of the zodiac subdivisions (Anastasiou et al., 2013; Bitsakis and Jones, 2016a).

Table 2: The duration of the astronomical seasons according to Democritus; Meton (observations in 432 BC); Euctemon (collaborated with Meton); Callippus (c. 370 BC – 300 BC); Geminus, work and parapegma collection (1st cent. BC); and Cleomedes (between the mid-1st century BC and 400 AD). The preserved numbers are in red. Spring has the maximum number of days, as the aphelion day occurs a little before the end of Spring. The Earth is located at perihelion close to the end of Autumn. Eudoxus' season durations are not in agreement with the solar anomaly.

| Astronomical Season/Period According to: | Winter Solstice up to 1 day before the Vernal Equinox | Vernal Equinox up to 1 day before Summer Solstice (includes the aphelion day) | Summer Solstice up to 1 day before the Autumn Equinox | Autumn Equinox up to 1 day before Winter Solstice (includes the perihelion day) | Duration of the Year |
|---|---|---|---|---|---|
| Meton | 92 | 93 | 90 | 90 | 365$^d$ |
| Euctemon | 92 | 93 | 90 | 90 | 365$^d$ |
| Callippus | 90 | <94> | 92 | 89 | 365$^d$ |
| Eudoxus | 91 | <91> | 91 | 92 | 365$^d$ |
| Democritus | 91 | [91] or [92] or [93] or [90] | [92] or [91] or [90] or [93] | 91 | 365$^d$ |
| Geminus (Elem. Astron.) | 90⅛ | 94½ | 92½ | 88⅛ | 365.25$^d$ |
| Geminus (Par. collection) | 89 | 95 | 92 | 89 | 365$^d$ |
| Cleomedes | 90¼ | 94½ | 92½ | 88 | 365.25$^d$ |

The 365 subdivisions of the zodiac dial ring correspond to the days for the solar tropical cycle, and at the same time they cannot be considered as 'days' for the lunar pointer: the number of days for one full turn of the lunar disc/pointer should be 27.321 subdivisions, as one full turn of the lunar pointer corresponds to one sidereal cycle[14] i.e. the lunar pointer after its rotation, returns to the same position in the same zodiac constellation[15] (the gearing of the Antikythera Mechanism agrees with this condition).

### 2.4 Four Scenarios for the Zodiac Ring Division

#### 2.4.1 Scenario 1

The zodiac dial ring was a time-measuring scale, just like the Egyptian dial calendar ring. The zodiac ring was divided into 365 equal subdivisions representing the rounded value of 365.25 days per one solar tropical year (Voulgaris et al., 2018a). The 365 equal distance subdivisions correspond to the 12 zodiac months and each zodiac month has an unequal number of days: the number of days per zodiac month depends on the number of days the Sun is found within the boundaries of the corresponding zodiac sign spanning 30° (see Table 2).

Zodiac months do not all have an equal number of days, due to the solar anomaly: the Sun around the perihelion needs fewer days to travel across a zodiac sign of 30° (higher angular velocity) and more days when it is around the aphelion (lower angular velocity). This means that the four seasons do not have the same duration.

The duration of the four seasons is mentioned by Democritus, Meton, Euctemon, Callippus, Eudoxus, Geminus and Cleomedes (see Blass, 1887; Bowen and Goldstein, 1988; Goldstein and Bowen, 1983; Hannah, 2002; Huxley, 1963; Lehoux, 2007; Manitius, 1880; Neugebauer, 1975; Rehm, 1913, 1941; 1949; van der Waerden, 1960; 1984a). Geminus and Cleomedes (Bowen and Todd, 2004; Manitius, 1898; Spandagos, 2002; Ziegler, 1891) state the duration of each astronomical season with an accuracy up to ⅛ of a day (Table 2). Probably they had measuring instruments with higher resolution than the older astronomers and a larger number of observations.

For practical measuring reasons, rounding the day fraction to an integer number of days is needed. This means that the rounding can be to the next or to the current day (e.g. 94½ can be rounded to 94 or to 95, as also 90⅛ is rounded to 90 or 91). Only about 20% of the zodiac dial ring is preserved, so we do not know the exact durations of the four seasons, nor the day distribution per zodiac month, except for the months ΧΗΛΑΙ and ΣΚΟΡΠΙΟΣ, which have 30 subdivisions/days (Bitsakis and Jones, 2016a).

The day distribution for each zodiac month can be calculated taking into account the number of days per astronomical season (Table 2). The two opposite zodiac months, which include aphelion (Gemini) and perihelion (Sagittarius), should have the maximum and the minimum number of days respectively. The distribution of days for the remaining zodiac months can be determined by adding a 'uniform' increment for every month when moving from Sagittarius towards Gemini, and a 'uniform' decrement for the months moving from Gemini towards Sagittarius (Neugebauer, 1975: 628, 1352, Figures 3 and 4; see also Voulgaris et al., 2018a: Figure 12).

Table 3: Analysis of the seasons and the durations of the zodiac months. The first row presents the duration of each season mentioned by Geminus. Following are five best rounding allocations to an integer number of days per zodiac month. For the calculations it was taken into account that the perihelion date belongs to the zodiac month of Sagittarius (with the shortest number of days) and the aphelion date belongs to the zodiac month of Gemini (with the largest number of days—see Voulgaris et al., 2018a). Note that five out of six options of the best rounding present Libra and Scorpio with 30 subdivisions-days, as are also preserved for Libra and Scorpio in Fragment C. Further below, the duration of the seasons mentioned by Cleomedes and the two best roundings in integer numbers are presented. The calendar of Dionysios is presented at the end, which was divided into 12 zodiac months: 11 months × 30 days + 1 month × 35 days or 36 days every four years (ΔΙΔΥΜΩΝΟΣ—Gemini; see Boeckh, 1863; van der Waerden, 1984b).

| Astronomical Season Duration According to Geminus | | | | | | | | | | | | |
|---|---|---|---|---|---|---|---|---|---|---|---|---|
| Number of days per Zodiac month (distribution) | Autumn 88$^{1/8}$ | | | Winter 90$^{1/8}$ | | | Spring 94$^{1/2}$ | | | Summer 92$^{1/2}$ | | |
| | Lib | Sco | Sgr Perihelion | Cap | Aqr | Psc | Ari | Tau | Gem Aphelion | Cnc | Leo | Vir |
| | 30 | 30 | 28$^{1/8}$ | 28$^{1/8}$ | 31 | 31 | 31 | 31$^{1/2}$ | 32 | 32 | 30$^{1/2}$ | 30 |
| Best rounding 1 | 30 | 30 | 29 | 29 | 30 | 31 | 31 | 31 | 32 | 32 | 30 | 30 |
| Seasons duration 1 | 89 | | | 90 | | | 94 | | | 92 | | |
| Best rounding 2 | 30 | 30 | 29 | 29 | 30 | 31 | 31 | 31 | 32 | 31 | 31 | 30 |
| Seasons duration 2 | 89 | | | 90 | | | 94 | | | 92 | | |
| Best rounding 3 | 30 | 30 | 29 | 29 | 30 | 30 | 31 | 31 | 32 | 32 | 31 | 30 |
| Seasons duration 3 | 89 | | | 89 | | | 94 | | | 93 | | |
| Best rounding 4 | 30 | 30 | 28 | 30 | 30 | 30 | 31 | 31 | 32 | 31 | 31 | 31 |
| Seasons duration 4 | 88 | | | 90 | | | 94 | | | 93 | | |
| Best rounding 5 | 30 | 30 | 28 | 29 | 30 | 31 | 31 | 31 | 32 | 31 | 31 | 31 |
| Seasons duration 5 | | 88 | | | 90 | | | 94 | | | 93 | |
| Best rounding 6 | 30 | 29 | 29 | 29 | 30 | 31 | 31 | 31 | 32 | 31 | 31 | 31 |
| Seasons duration 6 | 88 | | | 90 | | | 94 | | | 93 | | |
| Astronomical Season Duration According to Cleomedes | | | | | | | | | | | | |
| Seasons' Durations | Autumn 88 | | | Winter 90¼ | | | Spring 94½ | | | Summer 92½ | | |
| Best rounding A | 88 | | | 90 | | | 94 | | | 93 | | |
| Best rounding B | 88 | | | 91 | | | 94 | | | 92 | | |
| Best rounding C | 88 | | | 90 | | | 95 | | | 92 | | |
| Zodiac Month Duration of Dionysios' Calendar | | | | | | | | | | | | |
| Days/zodiac month | 30 | 30 | 30 | 30 | 30 | 30 | 30 | 30 | 35/(36) | 30 | 30 | 30 |
| Days per Season | 90 | | | 90 | | | 95/(96) | | | 90 | | |

Since there is a strong correlation between Geminus' astronomy and the Antikythera Mechanism (parapegma events on the plates, the information engraved on the Metonic spiral, intercalary months, excluded days and Saros spirals, the Exeligmos pointer, and the initial starting date of the Mechanism based on Geminus' definition for Exeligmos), the season duration mentioned by Geminus (Manitius, 1880: Chapter 1) was selected for the analysis in Table 3. That said, it is not clear if the season durations mentioned by Geminus were based on his own observations or on the work of earlier astronomers.

Based on Table 3, two out of the six rounded distributions are reconstructed and presented in Figures 8 and 9. By selecting the 'Best rounding 1' in Table 3, the 12 zodiac months were divided into days/subdivisions ranging from 29 to 32. The distribution of the number of days per zodiac month is symmetrical with respect to the line of Solstices. By applying the 'Best rounding 1' for the initial calibration date of the Antikythera Mechanism (23 December 178 BC—Winter Solstice—1st Day of Capricorn; Voulgaris et al., 2023a; 2023c), the dates for each zodiac month can be deduced (Table 4).

The phrase ΔΙΔΥΜΟΙ ΑΡΧΟΝΤΑΙ ΕΠΙΤΕΛΛΕΙΝ (Gemini begins to rise) engraved by the ancient craftsman on parapegma plate 1 (Bitsakis and Jones, 2016a), means that the Sun crosses the first boundary/entry of the Gemini sign and rises at the same time with it on this specific date.

For seven zodiac signs there is no prominent star in the sky on/near the first entry/boundary of the respective sign. Therefore, the position of these signs in the sky can only be found if the first entries have been predefined on specific dates, or if the first entries of these signs are prepositioned in the sky and therefore their corresponding dates can be calculated. An exception is the first entry in the Leo

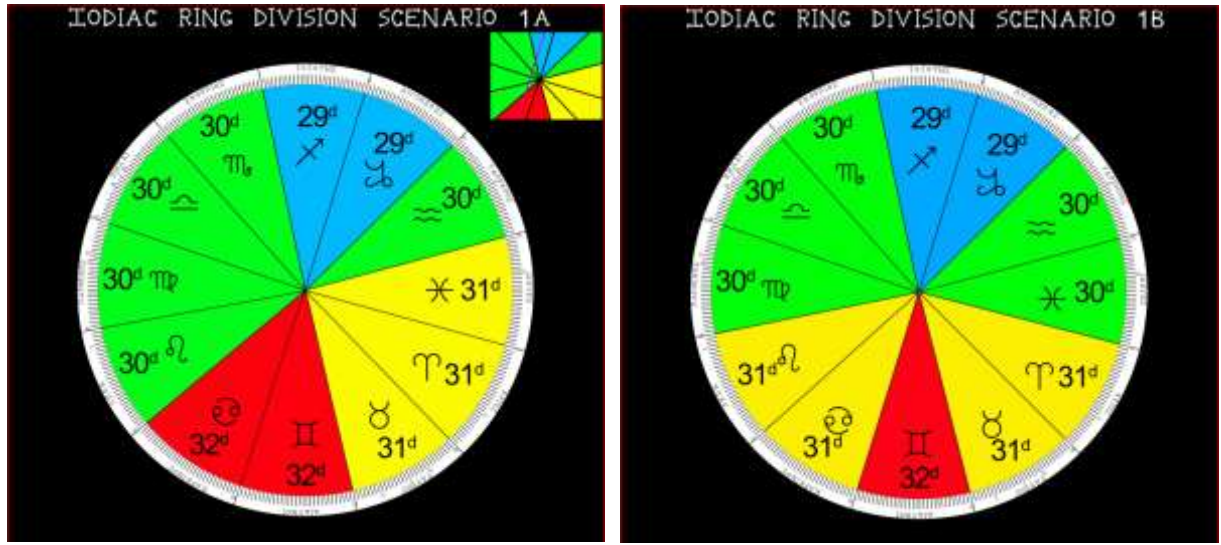

Figure 8 (left): An option for the zodiac dial ring divided into twelve zodiac months and a total of 365 equal subdivisions (days). The 12 zodiac months do not have the same numbers of days. For the zodiac ring reconstruction, the 'Best Rounding 1' of Table 4 was selected. Insert: the center of the zodiac ring. In magenta the line of Solstices connects the 1st of Capricorn and the 1st of Cancer. The line of Solstices does not cross the center of the zodiac dial ring due to the unequal lengths of the seasons. Note that no two 'opposite' radii form a real diameter. (About the headline: for the letter 'Z' the ancient craftsman used an 'I' with long serifs).

Figure 9 (right): A second option for the zodiac dial ring, divided into twelve zodiac months and a total of 365 equal subdivisions (days). The zodiac months do not have the same numbers of days. For the zodiac dial ring reconstruction, the 'Best Rounding 2' in Table 4 was selected. This option is also in agreement with the preserved 30 subdivisions of Claws (Libra) and Scorpion (after Bitsakis and Jones, 2016a).

Table 4: The duration of each zodiac month, based on Geminus' seasonal durations, using the 'Best Rounding 1' from Table 3. The calculated corresponding dates of the zodiac months start from 23 December 178 BC (1st day of Capricorn), which is the initial calibration date of the Antikythera Mechanism (Voulgaris et al., 2023a). The dates of perihelion and aphelion were calculated using the *Starry Night Pro* planetarium software.

| Zodiac Month | Zodiac Month Duration (Best rounding 1) | | Dates of Zodiac Months for 178 BC |
|---|---|---|---|
| Capricorn/Winter Solstice/ ΑΙΓΟΚΕΡΩΣ/ΤΡΟΠΑΙ ΧΕΙΜΕΡΙΝΑΙ | 29 | Winter Season Period: 90 days | 23 December–20 January |
| Aquarius/ΥΔΡΟΧΟΟΣ | 30 | | 21 January–19 February |
| Pisces/ΙΧΘΥΕΣ | 31 | | 20 February–22 March |
| Aries/Spring Equinox ΚΡΙΟΣ/ΙΣΗΜΕΡΙΑ ΕΑΡΙΝΗ | 31 | Spring Season Period: 94 days | 23 March–22 April |
| Taurus/ΤΑΥΡΟΣ | 31 | | 23 April–23 May |
| Gemini/ΔΙΔΥΜΟΙ | 32 | | 24 May–24 June (aphelion ~30 May) |
| Cancer/Summer Solstice ΚΑΡΚΙΝΟΣ/ΤΡΟΠΑΙ ΘΕΡΙΝΑΙ | 32 | Summer Season Period: 92 days | 25 June–26 July |
| Leo/ΛΕΩΝ | 30 | | 27 July (α Leonis/Regulus rises with the Sun on this date)–25 August |
| Virgo/ΠΑΡΘΕΝΟΣ | 30 | | 26 August–24 September |
| Claws/Libra/Autumnal Equinox ΧΗΛΑΙ/ΙΣΗΜΕΡΙΑ ΦΘΙΝΟΠΩΡΙΝΗ | **30** | Autumn Season Period: 89 days | 25 September–24 October |
| Scorpius/ΣΚΟΡΠΙΟΣ | **30** | | 25 October–23 November |
| Sagittarius/ΤΟΞΟΤΗΣ | 29 | | 24 November–22 December (perihelion ~29 November) |

sign, which presents a peculiarity: α Leonis/ Regulus (Basiliscus = 'The Little King' or 'The Lion Heart', named in the *Almagest*—see Heiberg, 1898–1903; see, also, Allen, 1963; Plakidis, 1992) is too close to the Ecliptic (see Voulgaris et al., 2023b: Figure 12), only ~21 arc min apart and rises with the Sun on 27 July in the era 180 BC (calculated with *Starry Night Pro* software) and this also applies for more than 150 years before and after 180 BC. There-fore, the zodiac sign of Leo could have a predefined rising date. The division of the Ecliptic into 12 sectors of 30° each can be easily achieved with high accuracy by starting the Ecliptic circle division with Regulus, using an astrolabe and an astronomical map.

### 2.4.2 Scenario 2

In the Eudoxus papyrus (Blass, 1887: Column I, 10–20) it is mentioned that the Sun travels from the South (Winter solstice) up to the North (Summer solstice) in 180 days and then stops its motion for two days (the Greek word Ηλιοστάσιο/Solstice means "… the Sun stops there."). The Sun travels from the North to the South in the same manner, for 180 days and then stops for three days in the South. The text continues, describing how from the Apeliotes/East (Vernal Equinox) to the North, the Sun travels for 90 days, from the Summer Solstice to the Livas (Lips)/SW (or WSW), though it should be Zephyros/West (Autumn Equinox), the Sun travels for 90 days, and for another 90 days from Livas to the South (see Table 5).[16] Therefore, the tropical year was divided either into 12 equal zodiac months of 30 days each, plus 3 days for the Winter Solstice plus 2 days for the Summer Solstice (Scenario 2A, after correction for the solar anomaly), or 30 + 2 = 32 days for Cancer, 30 + 3 = 33 days for Capricorn, and 30 days for the rest of the months (Scenario 2B after correction).

Table 5: According to Column I, 10–20 in the Eudoxus papyrus (Blass, 1887), the duration of each zodiac month is 30 days, but there are two extra days for the Summer solstice (before the 1st day of Cancer) and three extra days for the Winter solstice (before the 1st day of Capricorn). However, the specific ordinance of the extra days contradicts the solar anomaly effect. The present reference could be erroneous, since the opposite arrangement (here in red) of the extra days matches very well the solar anomaly/dates of perihelion and aphelion (the 3rd column in the table). The extra days could be considered as independent days (3rd column) or they could be included in the zodiac months of Cancer (33 days) and Capricorn (32 days) (the 4th column).

| Zodiac Month | Number of Days per Zodiac Month | Correction Based on the Solar Anomaly (Scenario 2A in Figure 10) | Probable Rounding Including the Correction (Scenario 2B in Figure 11) |
|---|---|---|---|
| Aries/Κριός | 30 | 30 | 30 |
| Taurus/Ταύρος | 30 | 30 | 30 |
| Gemini/Δίδυμοι | 30 | 30 | 30 |
| Summer Solstice/ΤΡΟΠΑΙ ΘΕΡΙΝΑΙ | 2(!) | 3 (engraved letters ΤΘ) | |
| Cancer/Καρκίνος | 30 | 30 | 33 |
| Leo/Λέων | 30 | 30 | 30 |
| Virgo/Παρθένος | 30 | 30 | 30 |
| Libra/Claws/Χηλαί | 30 | 30 | 30 |
| Scorpion/Σκορπιός | 30 | 30 | 30 |
| Sagittarius/Τοξότης | 30 | 30 | 30 |
| Winter Solstice/ΤΡΟΠΑΙ ΧΕΙΜΕΡΙΝΑΙ | 3(!) | 2 (engraved letters ΤΧ) | |
| Capricorn/Αιγόκερως | 30 | 30 | 32 |
| Aquarius/Υδροχόος | 30 | 30 | 30 |
| Pisces/Ιχθείς | 30 | 30 | 30 |
| Tropical Year duration | 365 days | | |

However, the values for the Winter and Summer Solstices days are not in accordance with the solar anomaly, so a correction is suggested (day values in reversed position)—see Table 5, columns 3 and 4.

Only ~20% of the zodiac month ring is preserved, and the decisive complete months of Sagittarius, Capricorn, Gemini and Cancer, and part of Leo are missing, which would allow us to draw some definitive conclusions about the view of the zodiac dial ring divisions, So we suggest two additional options—Scenario 2A and 2B—for the zodiac dial ring reconstruction. These are in accordance with the months that are fully preserved in Fragment C (see Figures 10 and 11).

## 3 HANDLING OF THE FRONT DIAL RINGS ON THE ANTIKYTHERA MECHANISM

When using the Antikythera Mechanism, a periodic correction procedure for the Egyptian calendar was needed: after four full rotations of the golden sphere (i.e. four solar tropical years), the Egyptian dial ring was rotated counterclockwise by one subdivision/day. Rotation of the Egyptian dial ring by mistake was probably avoided by using a small perpendicular pin with a head, which perforated the ring and was inserted in one of the 365 holes (Bitsakis and Jones, 2016a). The pin position could be right on the subdivision of the 1st of *Thoth*, which was the New Year of the Egyptian calendar or in the subdivision of the 18th of *Hathyr* (Voulgaris et al., 2023a; 2023c), which was the initial calibration date of the Mechanism (unfortunately, none of these months is preserved today). After four rotations of the golden sphere–Sun, the user pulled out the small pin, rotated the Egyptian dial ring counterclockwise by one subdivision, then inserted the small pin in a new hole. In this way, the Egyptian new year was always one day earlier than the (fourth) new (solar tropical) year (Bitsakis and Jones, 2016a; Price, 1974).

In Fragment C, the current position of the *Pachon* second subdivision corresponds to the 18th subdivision of *Chilai* (Libra) and this coincidence occurred around 571 BC (±4 years or ±8 years). A probable scenario is that a non-expert changed the positions of the two rings or an expert tried to calibrate the Mechanism

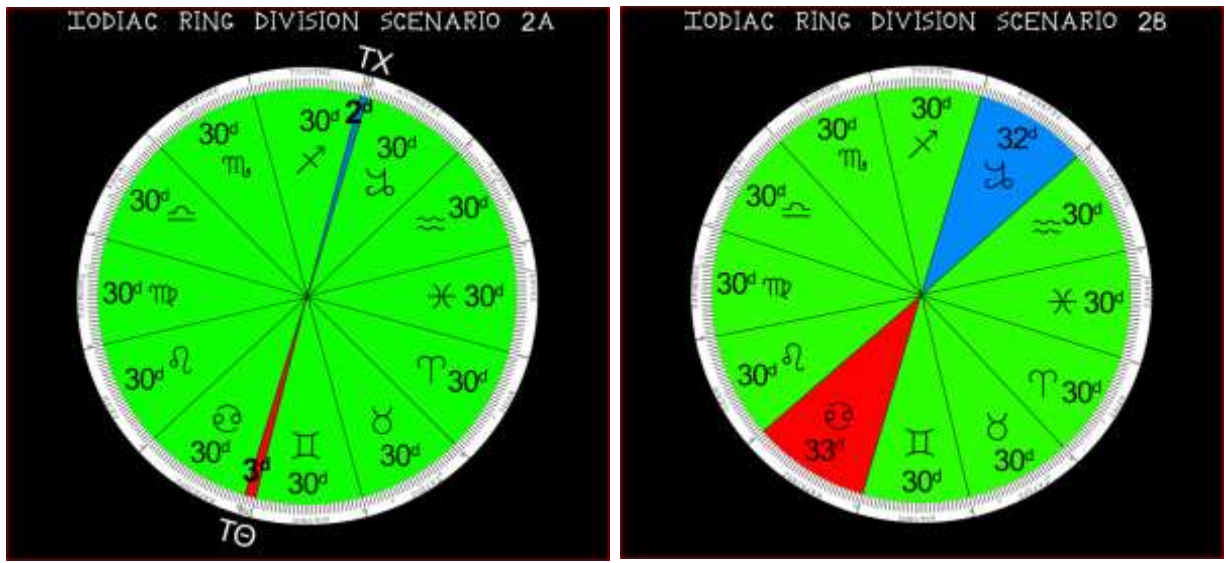

Figure 10: The first option (Scenario 2A) for the zodiac dial ring subdivision according to Eudoxus' papyrus (Blass, 1887). The zodiac dial ring is divided into 365 equal subdivisions/days, in 12 zodiac equal months of 30 days (and after the suggested correction on Table 5) plus 2 independent days for the Winter Solstice (with the engraved letters TX = ΤΡΟΠΑΙ ΧΕΙΜΕΡΙΝΑΙ) and 3 independent days for the Summer Solstice (with the letters TΘ = ΤΡΟΠΑΙ ΘΕΡΙΝΑΙ). This option is also in agreement with the preserved 30 subdivisions of the Claws (Libra) and the Scorpion (after Bitsakis and Jones, 2016a).

Figure 11: The second option (Scenario 2B) for the zodiac dial ring subdivision according to Eudoxus' papyrus (Blass, 1887). The zodiac dial ring is divided into 365 equal subdivisions/days, in 12 zodiac months: 33 days for Cancer, 32 days for Capricorn (after the suggested correction in Table 5) and 30 days for the rest of the months. This option is also in agreement with the preserved 30 subdivisions of the Claws (Libra) and the Scorpion (after Bitsakis and Jones, 2016a).

for a different date.

Before starting the time calculations, the pointers of the Mechanism should be calibrated according to the initial starting date of 23 December 178 BC, which corresponds to the 1st of Capricorn and to the 18th of *Hathyr* (Voulgaris et al., 2023a; 2023c)—see Figure 12.

During this unique date, four well-known lunar cycles were all at their starting position.[17] The Sun crossed the 1st entry of the zodiac sign of Capricorn (the Winter Solstice), there was an annular solar eclipse with a maximum duration (during BC years and right up to 150 AD), and the winter Metonic calendar and the 3rd Callippic winter cycle started. The Isia Feast was dedicated to Isis and Osiris, and also started on the 17th of *Hathyr* (Voulgaris et al., 2023a; 2023c). The fiducial line (Price, 1974; Voulgaris et al., 2023a) is engraved about 18 subdivisions clockwise from the middle vertical line on the square front plate[18] and coincides with the *VLb* (the vertical line that crosses the axis $b_{fn}$/center of the lunar cylinder and the centers of the Egyptian and the zodiac dial rings (see Voulgaris et al., 2019b; see, also, Allen et al., 2016). By rotating the Egyptian dial ring so that the 18th subdivision of *Hathyr* points at the fiducial line, the 1st of *Hathyr* crosses the middle vertical line of the square front plate (*VLb*—see Figure 11).

The ancient craftsman started calibrating his device by rotating the Egyptian dial ring, so that the 1st subdivision of *Hathyr* coincided with the *VLb* line. Afterwards, he engraved the fiducial line at the position of the 18th subdivision of *Hathyr*. Then, he rotated the zodiac dial ring so that the 1st subdivision of Capricorn pointed at the 18th subdivision of *Hathyr* and to the fiducial line. Then, the calibration of the golden sphere–Sun pointer and the lunar pointer and the Moon phases sphere followed (see Voulgaris et al., 2023a).

This peculiar/'clear' and hidden geometry on the positions of the zodiac dial ring and the fiducial line directly connects the Antikythera Mechanism with the date of 23 December 178 BC. On a time-measuring machine the random design and positioning of its parts, indicators, information, etc., can easily create misunderstandings, misconceptions and wrong conclusions (see Voulgaris et al., 2023c: Figure 5). It seems that the ancient craftsman placed every part carefully by adopting proper geometry and symmetry for the design, thereby avoiding random placements (Voulgaris et al., 2021).

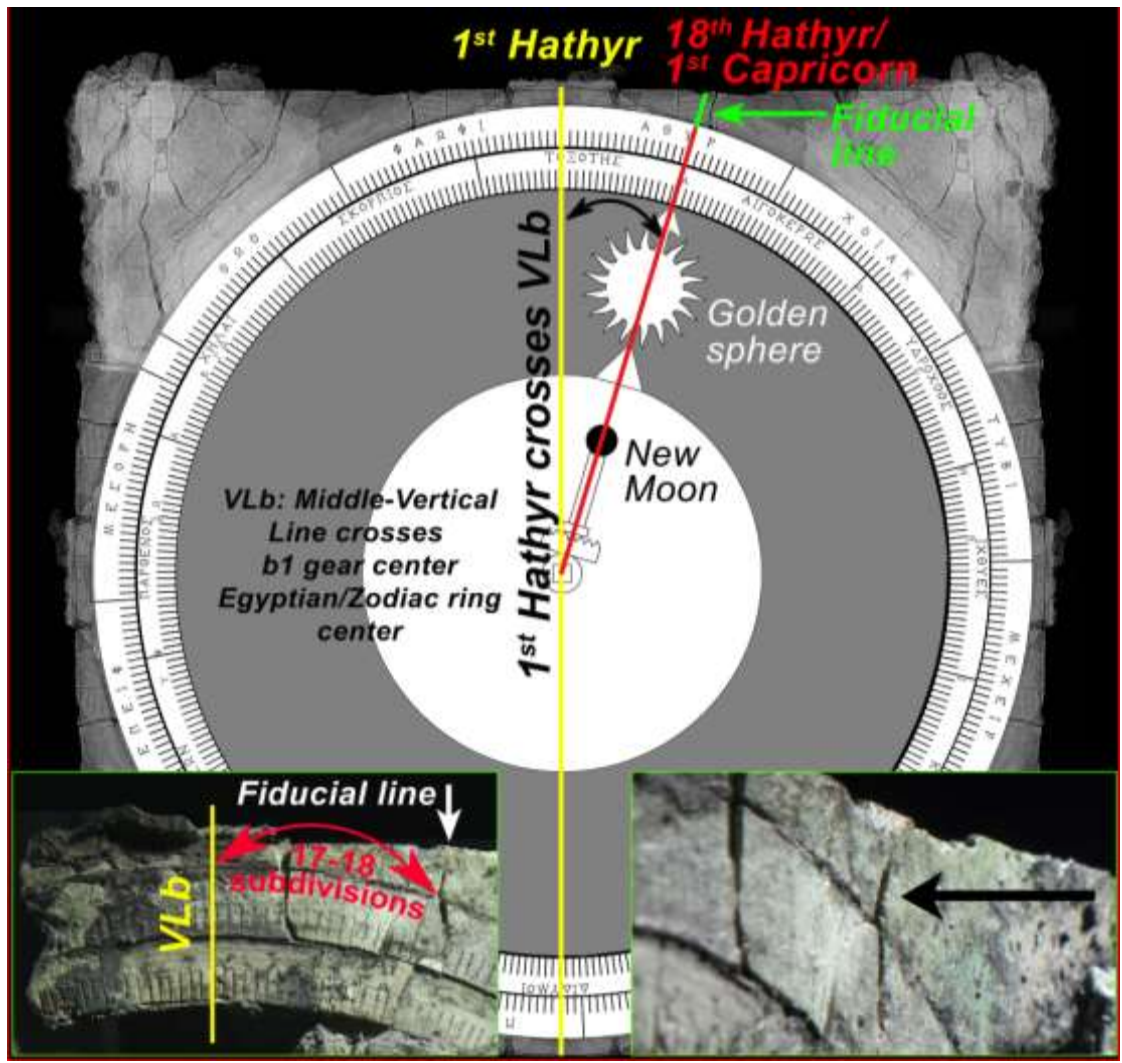

Figure 12: Calibrating the Egyptian and zodiac dial rings during the initial setting up of the Antikythera Mechanism. This AMRP radiograph of Fragment C, processed by the authors, was placed at one of the four corners (based on symmetry of the design), creating a digital reconstruction of the front square plate of the Mechanism. Note the remarkable geometry of the Antikythera Mechanism's front dial plate. The Inserts, left to right, show the fiducial line engraved at the 17th–18th subdivision to the right of the *VLb* line (in yellow). The Egyptian and Zodiac dial rings (via rotation) were calibrated by applying the initial starting date of the Mechanism's pointers, 23 December 178 BC (Voulgaris et al., 2023a), the 1st subdivision of Capricorn and the 18th subdivision of *Hathyr*, so they pointed directly to the fiducial line. Therefore, the 1st subdivision of *Hathyr* crosses the *VLb* line. The pointer of the golden sphere–Sun, ΗΛΙΟΥ ΑΚΤΙΝ, points to the fiducial line/1st of Capricorn and the lunar pointer points to the golden sphere–New Moon (and right after a new synodic month begins). The New Moon is also located at the the 1st of Capricorn (photograph credits: National Archaeological Museum, Athens, first author, Copyright Hellenic Ministry of Culture & Sports/Archaeological Receipts Fund).

## 4  CONCLUDING REMARKS: THE ANTIKYTHERA MECHANISM AS A TIME-MEASURING DEVICE

According to Tables 3–5 and the arguments presented in this paper, the zodiac dial ring co-operates better with the golden sphere–Sun and the Egyptian dial ring if its measurement units are units of time. It seems that the Antikythera Mechanism was a time-measuring device rather than a positioner, and it was constructed in order to offer time-calculation services: the dates and the hours of New Moon/ Full Moon and the star risings/settings; the hours of important future astronomical events, such as eclipses; and also the dates of the athletic Games.

During that era, time-keeping and event-prediction were of high interest, and were necessary and under continuous study by the Hellenistic astronomers. The Antikythera Mechanism offered a high capacity for time information, portability, and high repeatability with its calculations. Although geared machines presented inherent errors that affected the final positions of the pointers (Edmunds, 2011; Voulgaris et al., 2023b), this measuring machine was a unique ΧΡΟΝΟΥ ΑΥΤΟΜΑΤΟΝ time automaton (a time 'robot' of Antiquity, like Heron of Alexandria's ideas/constructions, see Schmidt, 1899) at that time, because it provided direct time information about events after only a few movements input and without additional complex, time-consuming calculations by the user.

This machine could replace the ancient 'time-keepers' who recorded the times and the events on a large number of papyri and clay tablets, but their management and storage was not efficient and too often these records could be lost or destroyed after a war or a famine or pestilence. The Antikythera Mechanism could be used by the Administrative Authority–Time Committee of that era.

As time measurement during that era was primarily based on lunar phases and secondly on the solar tropical year, only the relative (timed) positions of the two celestial bodies, the Moon and the Sun, were necessary for the calculation of time using the Antikythera Mechanism.

## 5 NOTES

1. Voulgaris et al. (2022) make a case for the existence of the missing Draconic cycle that should exist on the Antikythera Mechanism on the grounds of the Draconic gearing (as an engagement of gears b1–a1, r1/Fragment D and one hypothetical small gear s1; and revised in gears s1, s2, t1 in Voulgaris et al., 2024), which represent the fourth lunar cycle, well known in Antiquity. The eclipse events on the Saros spiral can be easily calculated based on a mechanical procedure and by only using the Mechanism's Draconic Pointer inside the Ecliptic limits and during Full Moon/New Moon (Voulgaris et al., 2023b, see also their Appendix 1: 29–64 and Voulgaris et al., 2024). The eclipse events engraved on the Saros spiral are not real recorded observations but rather end results of mechanical calculations done with the gears of the Mechanism.
2. The hours of eclipse events can be easily calculated by using only the Antikythera Mechanism and without any external information (unrelated to the Mechanism): by measuring the number of the zodiac ring subdivisions from New Moon/Full Moon to the next New/Full Moon, and converting them into days/hours (Voulgaris et al., 2023b: 22–25).
3. E.g. on Fragment C the lunar cylinder is stuck underneath the Egyptian/zodiac dial rings, parapegma plate 1 is stuck in front of the lunar cylinder, and one piece of parapegma plate 2 is stuck to the reverse face and in the opposite direction on parapegma plate 1. See also Figure 6 for a part of the front dial plate that is trapped inside the petrified silt of Fragment F.
4. AMRP radiography is Advanced Medical Radiography Practitioner radiography, and involves a senor radiographer with an advanced level of knowledge and skill.
5. Small parts of the parapegma plates with inscriptions are also preserved on Fragments 9, 20, 22, 28 (Bitsakis and Jones, 2016a).
6. On the back cover inscription, the text ΠΡΟΕΧΟΝ ΑΥΤΟΥ ΓΝΩΜΟΝΙΟΝ Σ[ΕΛΗΝΗΣ], and a (lunar) pointer projecting from it, are preserved (see Bitsakis and Jones, 2016b: 232).
7. ΗΛΙ[ΟΥ] ΑΚΤΙΝ, solar ray (see Bitsakis and Jones, 2016b: 233). The golden sphere –Sun should be linked to the b1 gear or the lost cover of the b1 gear (see Voulgaris et al., 2019b: Figure 1). The phrase 'solar ray' can be related to the solar pointer, which travels across the zodiac months and their subdivisions.
8. This date resulted from combining Geminus' definition of the exeligmos (Saros) cycle and the specific prominent and critical placement of the Winter Solstice inscription at the top-left corner of the Antikythera Mechanism (see Voulgaris et al., 2023c: Figure 5). Other researchers followed different approaches and they arrived at different starting dates, see Carman and Evans, 2014; Freeth, 2014; Iversen, 2017; Jones, 2020.
9. The axis of the b1 gear is located at the center of the front plate of the Mechanism (e.g. at the intersection of the two diagonals of the rectangle shape of the Mechanism), which is the center of the lunar cylinder and the common center of the Egyptian and zodiac dial rings. This leads to the conclusion that the two parapegma plates should have had exactly the same dimensions.
10. The specific number of 365 holes is mandatory for the proper functioning of the Egyptian dial ring (see Section 3). A different number of holes would cause the Mechanism's measurements and operation to malfunctions.
11. Freeth and Jones (2012) suggest that this sliding catch was part of the back-cover plate. Wright (2005: Figures 10 and 11; 2011: 12) suggests that it was a part of the back plate, and therefore the back plate could be detached from the main box of the Mechanism (like the detachable central square plate). However, at the right boundary of the back plate (Fragment A2) three long perpendicular pins are preserved that perforate the back plate and a piece of wood fixing the bronze back plate to the internal wooden casement. Therefore, the back plate and internal wooden casement were one fixed piece/casement, and all of the other parts of the Mechanism's body (the middle plate, wooden spacers, etc.) were nested inside this casement (see Voulgaris et al., 2019b: Figures 6 and 7).

12. The position of the draconic pointer on the Antikythera Mechanism represents the position of the Moon in the zodiac belt. When the draconic pointer lies between the Ecliptic limits a solar or a lunar eclipse occurs (depending on the relative position of the lunar pointer/Moon to the golden sphere–Sun). When the draconic pointer crosses the Ecliptic line, a central eclipse will occur (see Voulgaris et al., 2022: Figure 17 and Voulgaris et al., 2023b: Figure 15). Meanwhile, for information about Geminus see Evans and Berggren (2006).

13. Therefore, 365.25 days/365 subdivisions ≈ 1.0006849 days/subdivision (≈ $24^h$ $0.986^m$), and this difference is practically undetectable when taking into account minor division errors, mechanical mismatches, and the width of the engraving tool's nose.

14. During one full turn of the lunar pointer = one sidereal cycle of 27.321675 days ≈ 655.72 hours (according Geminus' description of exeligmos and lunar cycles in Chapter XVIII; see, also, Voulgaris et al., 2023b), the lunar pointer transited the 365 subdivisions of the zodiac dial ring. Therefore, 655.72 hours/365 subdivisions ≈ 1.8 hours/subdivision. This is the answer to the question "How did the ancient craftsman know when eclipses would occur?" (some solar eclipses occurred at night). Note that any external information unrelated to the Antikythera Mechanism about the hour of the eclipse could differ from the value calculated with the Mechanism since all of the geared devices contained positioning and timing errors. However, if the Mechanism could be used to calculate the contact times of these events (see Voulgaris et al. 2023b: 22), then there was no reason for an ancient craftsman to use external, unrelated information to determine these eclipse times.

15. Each pointer was accompanied by its corresponding calibrated scale; the lunar disc probably had its 'personal' scale of 29.5 day-subdivisions based on the synodic cycle engraved on the (lost) cover of b1 gear (see Figure 7). A large number of astronomical clocks of the Medieval period present a circular lunar age scale divided into 29.5 subdivisions/days (see Guilbaud, n.d.).

16. For the directions of the winds see Anastasiou et al., 2016: 205; Wood and Symons, 1894: 81, Figures 1 and 2).

17. According to Geminus, the Exeligmos (Saros) starts with the apokatastasis of the (three) lunar cycles, i.e. when the synodic, anomalistic and draconic cycles are all located at their beginnings (Voulgaris et al., 2023a: 8–16).

18. The preserved area of the square front plate includes its middle-vertical line which coincides with *VLb*.

## 6   ACKNOWLEDGEMENTS

We are very grateful to Professors M. Edmunds, J. Seiradakis and X. Moussas who provided us with the license permission and the X-ray raw volume data of the Antikythera Mechanism fragments. Many thanks to Dr. F. Ullah for his support in our use of the REAL3D VolViCon software.

Finally, thanks are due to the National Archaeological Museum of Athens, Greece, for permitting us to photograph the fragments of the Antikythera Mechanism.

# 8 APPENDIX

## 8.1 Dividing the Zodiac Dial Ring into Units of Time Versus Units of Position–Space

The Antikythera Mechanism is a geared measuring instrument. By means of its gears, scales and pointers it calculates the dates of astronomical and social events. Each measurement/ result is presented with the help of a rotating pointer which has a calibrated scale. A calibrated scale is a scale which is divided into a number of subdivisions—parts of a circle (or parts of spiral). Each of the subdivisions is the minimum possible measurement and it presents the minimum measuring unit of the scale (Voulgaris et al., 2022: 108, and Figure 3).

The nature of the units in a measuring instrument defines the kind of instrument. Measuring instruments have calibrated scales that present relevant or interrelated measurements. For example, it is difficult to find a device that measures the intensity on sound (in dB) and the optical activity of carbohydrates (in arc degrees), as there is no clear correlation between the units of measurement.

The main idea of this paper is the advantages (functionality and uniformity) that result from our suggestion that the Antikythera Mechanism zodiac dial ring was divided into 365 equal subdivisions (days, units of time) instead of 360 (equal or unequal) subdivisions (degrees, or units of space), as was suggested by Bitsakis and Jones (2016a); Evans et al. (2010); Freeth et al. (2006); Price (1974) and Wright (2006; 2007; 2012).

By dividing the zodiac dial ring into 360 equal degrees, the solar anomaly does not have to be represented on the Antikythera Mechanism.

Evans et al. (2010) suggest that the zodiac ring should be divided into 360 unequal period subdivisions in order to represent the solar anomaly: denser divisions for zodiac signs close to/around Sagittarius, since the date of perihelion (fastest solar angular velocity) is around 29 November (and the 1st of Capricorn Winter Solstice on 23 December for ca. 180 BC) and more expanded divisions for zodiac signs close to or around Cancer, since the date of aphelion (with the slowest solar angular velocity) is around 30 May and the 1st of Cancer is on 26 June (the Summer Solstice for ca.180 BC), dates that were calculated with the planetarium software *Starry Night Pro*.

The Egyptian dial ring has a common center with the zodiac dial ring, and presents equal design and equal operation, since it measures the Egyptian year, having 365 equal subdivisions (i.e. days).

In the same manner, the zodiac dial ring could also be a time-measuring scale representing the duration of the solar tropical year, also divided into 365 equal subdivisions (again, days). The two ring scales are traversed by the solar pointer of the golden sphere–Sun. With these three parts, the measurement of the year is achieved (see below).

The 365 subdivisions of the zodiac dial ring cannot be considered as days for the lunar pointer, as one full turn of the lunar pointer corresponds to one sidereal lunar month of 27.321 days. The 365 subdivisions can be considered as hour units during the lunar path: 27.321 days/365 subdivisions ≈ 1.8 hours/subdivision (see Voulgaris et al., 2023b).

The lunar disc/lunar pointer could also have had its personal scale, the lunar age scale divided into 29.5 subdivisions, since a day scale for the lunar synodic cycle is missing from the Antikythera Mechanism.

## 8.2 The Solar and the Lunar Pointers of the Antikythera Mechanism

Each calibrated measuring scale is accompanied by its corresponding pointer. Let us analyze the pointers on the Antikythera Mechanism.

On the back cover inscription (Bitsakis and Jones, 2016b) the 'Instruction Manual' of the Antikythera Mechanism is partially preserved. In this text, the ancient craftsman presents the operational (not the decorative) parts of the instrument. In line 16 the lunar pointer is presented: ΠΡΟΕΧΟΝ ΑΥΤΟΥ ΓΝΩΜΟΝΙΟΝ Σ[ΕΛΗΝΗΣ [projecting from (the lunar disc) a (lunar) pointer]. The main operation of this pointer should be its position in relation to the Sun:

- Lunar pointer points to the Sun (New Moon),

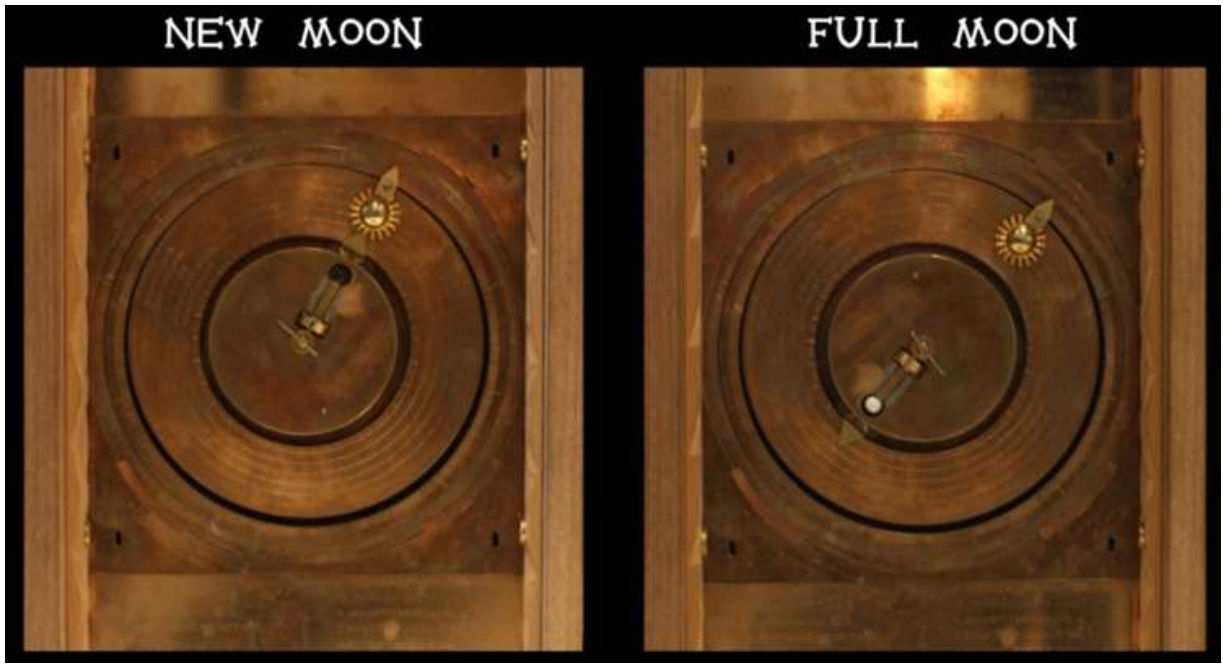

Figure 13: Close up on the square front plate of the authors' functioning reconstruction of the Antikythera Mechanism The two characteristic lunar phases, New Moon and Full Moon (time difference of half a synodic month) are presented. Note that the golden sphere–Sun has also traveled through about 15 zodiac and Egyptian subdivisions. The bronze surfaces are suffering from (atmospheric) humidity oxidation and the bronze bright light is lost due to copper oxide in the surface (after Voulgaris et al., 2019b).

- Lunar pointer in the direction away from the Sun (Full Moon).

These two positions were the most critical for the ancient Hellenic Metonic calendars, and for the Antikythera Mechanism (dates-month beginning, eclipses, Games) and of lesser importance the 7-day and 21-day lunar phases, lunar-solar pointers at 90°:

- The New Moon phase marked the end of the previous month, the beginning of a new month, and the possibility of a solar eclipse.
- The Full Moon phase marked the middle of the month, and the probability of a lunar Eclipse.

In addition, the prediction of the New Moon /Full Moon phase for a future date gave additional information: the dates around Full Moon would have bright nights, allowing night-time activities such as traveling, hunting, navigation, military operations, etc. Figure 13 shows a close-up of the front plate of the Antikythera Mechanism with the New Moon and Full Moon phases activated.

On the Antikythera Mechanism's back cover inscription line 21 says: ΓΝΩΜΩΝ ΚΕΙΤΑΙ ΧΡΥΣΟΥΝ ΣΦΑΙΡΙΟΝ (the pointer lies on the golden sphere) and in Line 22 ΗΛΙΟΥ ΑΚΤΙΝ (Sun ray; Bitsakis and Jones 2016b, see also https://arxiv.org/pdf/2207.12009). The preserved words are directly related to the

Golden sphere–Sun (Figure 7) and its pointer, the Sun ray, which is fixed on the golden sphere (see Figure 14). The operation of the golden sphere and its pointer are related to the zodiac and Egyptian ring scales, as one full turn of the golden sphere/pointer corresponds to a solar tropical year. Additionally, when the Sun ray pointer points at a zodiac subdivision with an index letter, then at this date a stellar event occurs (which is engraved on one of the two parapegma plates).

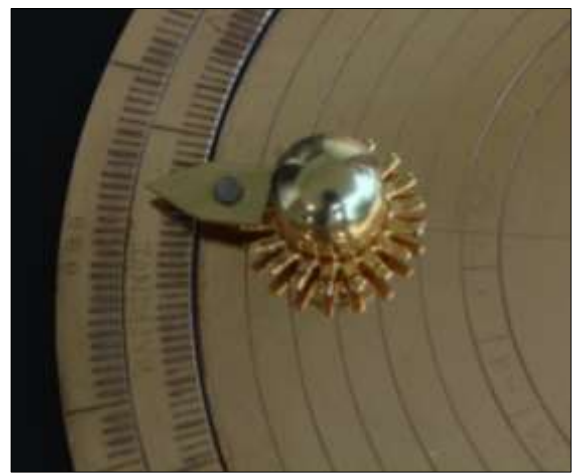

Figure 14: A close-up of the solar ray and the zodiac and Egyptian ring subdivisions on a reconstructed copy of the Antikythera Mechanism. The ΗΛΙΟΥ ΑΚΤΙΝ points to the 27th day of Cancer and at the same time it points to the 20th day of *Thoth* (photograph: the authors).

Table 6: Protocol of Mandatory Parameters and Scientific Methodology for a Research Quality Bronze Reconstruction of the Antikythera Mechanism.

| | Mandatory Parameters |
|---|---|
| 1 | Must be based on CT scans and photographs of the fragments. |
| 2 | Without any change of the preserved parts' dimensions and ordered position (taking into account the symmetry of the design, and also the deformation and the displacement of the parts). |
| 3 | Without making holes where none existed on the ancient original prototype. |
| 4 | A hypothetical part cannot be adapted in a place where there is no indication of a missing component. |
| 5 | Keeping/repeating the *Personal Constructional Characteristics* of the ancient craftsman. |
| 6 | Without using any modern mechanical parts (screws, bolts, grub screws, etc.). |
| 7 | Having a gear thickness of around 2mm and without hubs (but with a central increment in the thickness). |
| 8 | The teeth of the gears must be triangular in shape (not a modern involute shape). |
| 9 | Using a simple bronze alloy for the bronze gears, bronze shafts and bronze axes (and not a special bronze alloy or steel or other ferrous alloys). |
| 10 | Using the minimum hypothetical parts' addition. |
| | Scientific Methodology |
| a | Hypothetical considerations (e.g. units of a lost scale) are approved only after directly checking the consequences, impacts and functionality of an hypothesis against the results, and are rejected only by the use of a functioning research-quality model of the Antikythera Mechanism. |
| b | Any hypothetical operation of the Antikythera Mechanism must be supported by an experiment or by a bronze functioning reconstruction. |
| c | If an hypothesis disagrees with the experiment it is wrong. |
| d | Theoretical approaches that are confirmed non-experimentally cannot be accepted *ipso facto*, and they cannot be used to reject other hypotheses that are supported by experiments. |

## 8.3 Zodiac Ring Division into 365 Days

Based on the preserved 30 subdivisions of Libra and the 30 subdivisions of Scorpio on Fragment C, Price (1974) argued that the zodiac ring is divided into 12 × 30 = 360 subdivisions/degrees.

But according to Table 3, five of the six 'Best rounding' options also presented Libra and Scorpio with 30 subdivisions/days. Therefore, the preserved data cannot support the hypothesis that the zodiac ring was divided into 360 degrees. Unfortunately, the critical zodiac sectors that could define the final units of the zodiac ring are Sagittarius/Capricorn and Taurus/Gemini/Cancer/Virgo, all of which are lost today.

In our suggestion, the solar anomaly is well represented and as the parapegma events occur on specific dates (not in degrees of ecliptic longitude), the parapegma units/days are in agreement with the zodiac ring subdivisions/days (the index letters are engraved on some of the zodiac subdivisions; see Bitsakis and Jones 2016a).

The division of the two rings into 365 days creates a visualization of a time difference between the two calendars, as Geminus also mentions. By observing the two ring scales the user can find which Egyptian day corresponds to a zodiac day (and *vice versa*, see Figure 14). The solar ray aims simultaneously at a zodiac ring day and its corresponding Egyptian day (for the specific year).

## 8.4 Construction Protocol for a Research Quality Bronze Reconstruction of the Antikythera Mechanism

After processing the AMRP X-ray raw data, re-orienting the tomographies (using the proper software) and examining the photographs of the fragments, we measured the calibrated data and digitally redesigned and completed the fragments of the Antikythera Mechanism.

In order to replicate as closely as possible the original artifact, we applied a specific 'Construction Protocol' of 10 + 4 parameters, which would lead to a research quality model reconstruction of the Antikythera Mechanism.

Based on the protocols listed in Table 6 we reconstructed three functioning models of the Antikythera Mechanism in bronze (see Voulgaris et al., n.d.(a); n.d.(b).

Note that any deviation from the 14 parameters could strongly alter the mechanical status of the Antikythera Mechanism system, and lead to different mechanical behavior, different results and conclusions and finally to incorrect considerations.

One of the most characteristic examples of such a mechanical status change after adapting different parts was the almost fatal aeroplane disaster on 10 June 1990, after the use of wrong-sized bolts (2.5 mm shorter than the correct ones!) were used in the pilot's window (https://www.newscientist.com/article/mg13418180-300-wrong-bolts-sent-pilot-into-the-blue/).

Any hypotheses about the operation of the Antikythera Mechanism can be tested using the experimental bronze reconstructions. If an hypothesis disagrees with the experiment it is wrong (see Feynman, 1964).

Our functional models have helped us to reconstruct and understand the behavior of the original Antikythera Mechanism, its mechanical limits and errors—which are inherent problems in all geared machines, even modern ones (see Voulgaris et al., 2023b, 12–20). For example, the transmission motion via gears with triangular teeth is 'broken'/intermittent/non-continuous (while modern design gears exhibit much smoother motion (see Voulgaris et al., 2023b).

These mechanical characteristics, and also the limited torques and friction (see Roumeliotis 2018) cannot be detected or represented in 3D digital representations, because software is written to operate as the manufacturer defines it.

### 8.5 The Hypothesis of the Planetary Indication Gearing in the Antikythera Mechanism

The (hypothetical) existence of the planets on the Antikythera Mechanism affects the units of the zodiac dial ring. The hypothesis of the planets' indication gearing presents a large number of mechanical and operational problems. A number of Antikythera Mechanism models present planet-pointers on their front cadran. On some of these models we detected a number of strong contradictions from our *Protocol for a Research Quality Bronze Reconstruction of the Antikythera Mechanism*, as they deviated from the original ancient prototype and on some other models, modern parts and modifications for their stabilization had been introduced in these constructions. Many suggested designs for the planets on the Antikythera Mechanism were not in agreement with the fourteen parameters in our protocol list (Table 6, above).

Although the names of the planets or their theophoric names are (well, partially or poorly) referred to on the back cover inscription on the Antikythera Mechanism (Bitsakis and Jones 2016b: 232–233), a more careful and attentive reading indicates that the planets' names were directly related to the word ΚΥΚΛΟΣ (circle-orbit), such as ΕΣΤΙΝ Ο ΤΟΥ ΚΡΟΝΟΥ ΦΑΙΝΟΝΤΟΣ ΚΥΚΛΟΣ—there is the circle of Kronos Phainon (but no ΕΣΤΙΝ ΤΟΥ ΚΡΟΝΟΥ ΦΑΙΝΟΝΤΟΣ ΤΟ ΣΦΑΙΡΙΟΝ).

In the same manner of syntax we have:
ΚΥΚΛΟΣ ΕΣΤΙΝ Ο ΤΟΥ ΑΡΕΩΣ ΠΥΡΟΕΝΤΟΣ (Circle of Mars),
ΚΥΚΛΟΣ ΕΣΤΙΝ Ο ΤΟΥ ΔΙΟΣ ΦΑΕΘΟΝΤΟΣ (Circle of Jupiter).

The ancient craftsman used about 2++ lines of the text (lines 20–22), in order to describe the Sun of the Mechanism:

- Its sphere and its color;
- Its position;
- Its pointer;
- Its operation; and also
- To start the description of which component will be presented next.

For the planets Mars (line 23), Jupiter (line 24) and Saturn (line 25) he spent only one line on each of these. In that one line and in just 70–87 letters he described the planet's sphere, its color, its position, its pointer and its operation, and he also mentioned which component would be presented on the next line. So, why did he not use more lines for each planet, as he did in the case of the Sun (the golden sphere)? An extensive analysis and reconstruction of the Antikythera Mechanism *Instruction Manual* is presented in the following link: https://arxiv.org/abs/2207.12009.

Additional arguments and answers regarding extremely doubtful hypotheses about the Antikythera Mechanism and planets are presented at https://arxiv.org/abs/2407.15858. The hypothesis is that the planets' indication gearing in the Antikythera Mechanism is not in accordance with the *Protocol of Mandatory Parameters for a Research Quality Bronze Reconstruction of Antikythera Mechanism.* There are many mechanical problems, and the preserved text of the back cover inscription does not prove the existence of the planets' spheres.

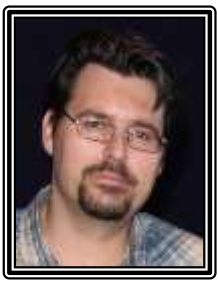

**Aristeidis Voulgaris** designs and constructs optomechanical and astronomical instruments. Since 2003 he has been an external partner of the Optics and Spectroscopy Laboratory in the Physics Department at the Aristotle University of Thessaloniki.

He has a research interest in spectral analysis of the corona during total solar eclipses, and has participated in 13 solar eclipse expeditions in collaboration with the late Professor Jay M. Pasachoff. One of Aristeidis' spectrographs flew on a NASA aircraft during the total solar eclipse of 2017, and he and his spectrographs also were on special flights to observe the total solar eclipses of 2019 and 2023 from close to Easter Island and Antarctica, respectively.

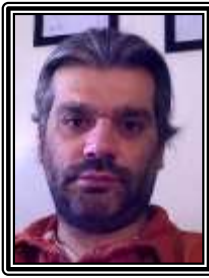

Aristeidis' other passion is to study the Antikythera Mechanism, and he is the team leader of 'The Functional Reconstruction of Antikythera Mechanism – The FRAMe Project'. He and his team have constructed several functioning bronze models of the Antikythera Mechanism, using the lathe and drilling milling machines in his own machine shop.

**Dr. Christophoros Mouratidis** has a PhD in mathematics. He has been working for about twenty years in various academic institutions, and at present he is an Assistant Professor at the Merchant Marine Academy of Syros in Greece.

His research interests are in complex analysis, harmonic analysis, and discrete mathematics. He also analyzes the spectra from the solar corona, which were taken during total solar eclipses.

He is a member of The Functional Reconstruction of Antikythera Mechanism – The FRAMe Project.

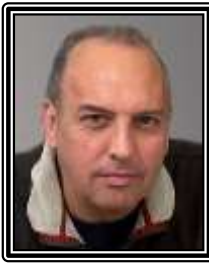

**Andreas Vossinakis** is an industrial designer, and his interests are in astronomical map designs, graphics, and tomography processes and analysis.

He is also a science communicator, especially in Astronomy. He is a member of The Functional Reconstruction of Antikythera Mechanism – The FRAMe Project.